# First-principle study of octahedral tilting and Ferroelectric like transition in metallic LiOsO$_3$

Hyunsu Sim and Bog G. Kim[*]

*Department of Physics, Pusan National University, Pusan, 609-735, South Korea*

The octahedral tilting and ferroelectric-like structural transition of LiOsO$_3$ metallic perovskite [Nature Materials **12**, 1024 (2013)] was examined using first-principles density-functional theory. In LiOsO$_3$, $a^-a^-a^-$ octahedral titling mode is responsible for the cubic ($Fm\bar{3}m$) to rhombohedral ($R\bar{3}c$) structural transition, which is stable phase at room temperature. At low temperatures, a non-centrosymmetric transition to a $R3c$ rhombohedra phase was realized due to zone center phonon softening. The phase transition behavior of LiOsO$_3$ can be explained fully by density functional calculations and phonon calculations. The electronic structure and Fermi surface changes due to the electron lattice coupling effect are also presented. The carrier density of state across the phase transition is associated with the resistivity, heat capacity, and susceptibility.

Recently, there has been considerable interest in octahedral tilting transitions and non-centrosymmetric transitions in perovskite oxides not only for their technological importance in ferroelectric materials and multistage memory devices but also for their scientific importance [1-4]. In this pursuit, first-principles density functional theory has played an important role not only in explaining the physical state but also in predicting the physical state of materials [5-7]. In an $ABO_3$ perovskite oxide material, the B ion forms an octahedron with neighboring oxygen atoms and the $BO_6$ octahedron is linked in 3 dimensional networks by sharing the corner oxygen atoms. Many physical states can be realized in the perovskite oxides depending on the size and ionic character of the A site atom and B site atom.

$LiOsO_3$ is one of the important and unique additions to oxide materials with a perovskite structure. According to Y. Shi *et. al.* [8], $LiOsO_3$ is a perovskite oxide material with metallic properties. Interestingly, at ~140 K, a centrosymmetric ($R\bar{3}c$) phase to non-centrosymmetric ($R3c$) phase transition was observed after careful examinations of the neutron scattering data and convergent-beam electron diffraction data. They also reported an anomaly in the temperature dependence of heat capacity, susceptibility and resistivity. They also indicated that such a structural phase transition is equivalent to the ferroelectric transition of $LiNbO_3$ materials.

In both centrosymmetric ($R\bar{3}c$) and non-centrosymmetric ($R3c$) phases, octahedral tilting is one of the important factors [9-12]. The three dimensional network of $OsO_6$ octahedral tilting in $LiOsO_3$ can be characterized easily by the Glazer notation [9]. The Glazer notation describes the octahedral tilting using the symbol $a^{\#}b^{\#}c^{\#}$, in which the literals refer to the tilt around the [100], [010], and [001] directions of the cubic perovskite, and the superscript # takes the value 0, +, or – to indicate no tilt or tilts of successive octahedra in the same or opposite sense. The rhombohedral ($R\bar{3}c$ and $R3c$) structure of $LiOsO_3$ can be classified by $a^-a^-a^-$ tilting [8, 9]. This tilting of $LiOsO_3$ is basically the same as $LiNbO_3$ [13-18], and originated from the rigidity of an $OsO_6$ octahedron and large ionic size difference of Li and Os ions.

In addition to octahedral tilting, the zone center soft mode is the main source of centro-symmetric ($R\bar{3}c$) and non-centrosymmetric ($R3c$) phases. In $LiNbO_3$, the phase transition from a centro-symmetric ($R\bar{3}c$) to non-centrosymmetric ($R3c$) phase occurs around 1480 K [1, 13, 18]. This phase transition is closely related to the softening of the $A_{2u}$ infrared active phonon mode, which is involved in the movement of Li ions and off-centering of Nb ions [13-18]. K. Parinski, Z. Q. Li, and Y. Kawazoe [13] applied first- principles density functional calculations to explain the phase transition in $LiNbO_3$ materials. Q. Ping and R. E. Cohen [18] used molecular dynamics with a first-principle-based potential to explain the temperature dependence of the pyroelectric properties and spontaneous polarization in $LiNbO_3$.

Here we report the detailed analysis of the octahedral tilting and non-centrosymmetric phase transition of LiOsO$_3$ materials. The structure was optimized using first-principles calculations and then, quantitative analysis was applied to octahedral tilting. The phonon dispersion curve was then calculated by frozen phonon calculations [13-15]. The analysis showed that the soft A$_{2u}$ mode at the zone center is responsible for the non-centrosymmetric transition in LiOsO$_3$ and the importance of the spontaneous breaking of symmetry in this material. Finally, the electronic structure and Fermi surface topology of each phase and their relationship were examined.

First-principles calculations were performed using the Perdew-Burke-Ernzerhof generalized gradient approximation (PBE GGA) [19] to the density functional theory and the projector-augmented-wave method, as implemented in VASP [20, 21]. The following valence electron configuration was considered: $1s^22s^1$ for Li, $5d^76s^1$ for Os, and $2s^22p^4$ for O. The electronic wave functions were expanded with the plane waves up to a kinetic-energy cutoff of 400 eV except for structural optimization, where a kinetic energy cutoff of 500 eV was applied to reduce the effects of Pulray stress. Momentum space integration was performed using a $12 \times 12 \times 12$ Gamma-centered Monkhorst-Pack k-point mesh [22]. With the given symmetry of perovskite imposed, the lattice constants and internal coordinates were optimized fully until the residual Hellmann-Feyman forces became smaller than $10^{-4}$ eV/Å. The calculated total energy as a function of the fixed volume was used to obtain the equation of states. At each given volume, the cell parameters and internal atomic coordinates were fully optimized. To obtain the phonon dispersion curve and phonon partial density of state, the frozen phonon calculation was applied to a $2 \times 2 \times 2$ supercell (containing 80 atoms in superlattice configuration) using the phonopy program [23]. The ISOTROPY and AMPLIMODES programs were utilized to check the group subgroup relationship and quantify the octahedral tilting as well as soft mode [24, 25]. To calculate the Fermi surface topology, the Wannier interpolation technique was applied using the WANNIER90 program [26].

Fig. 1(a) shows the rhombohedral phase of LiOsO$_3$. Without any octahedral tilting [9], LiOsO$_3$ has a P$m\overline{3}m$ cubic structure. With a$^-$a$^-$a$^-$ tilting [8,9], the $R\overline{3}c$ rhombohedral phase is stabilized and the unit cell can be viewed as [a, a, 0], [a, 0, a], and [0, a, a] lattice vectors, where a is the lattice constant of the pseudo cubic cell. This pattern can be viewed from the [111] axis of the pseudo cubic structure. Fig. 1(b) and 1(c) show a schematic diagram of two rhombohedral phases along the [111] axis of pseudo cubic lattice. In the centrosymmetric $R\overline{3}c$ rhombohedral phase, an Os atom is located at the symmetric position from two oxygen planes, which is also the central point connecting two Li atoms. A triple rotation with mirror symmetry ($\overline{3}$) is due to the oxygen octahedral. Below the phase transition temperature, the mirror symmetry is broken ($R3c$) because Os moves along the [111] axis, and the two oxygen planes are no longer equivalent to Os. In addition, the shift of small Li atom occurs simultaneously. As a result, the distance from Os atom to the neighboring oxygen atoms as well as that to Li atoms is not equivalent. The energy per formula unit vs. the volume per formula unit was calculated for three phases in Fig. 1(d). The lowest energy state is the non-centrosymmetric phase of $R3c$, where the total energy of $R\bar{3}c$ was 25 meV higher than that of $R3c$, and the total energy of the cubic phase was 978 meV higher than that of $R3c$. Table 1 lists the lattice parameters and Wyckoff position of each atom. The lattice parameters were slightly larger than that of the experimental values [8], which is common for a GGA approximation. Therefore, the optimized structure reproduced quite well the experimental values of the atomic position and lattice parameters.

To determine if these phase transitions are spontaneous, the total energy of system was calculated as a function of the mode amplitude. Each mode related to the phase transition was calculated with ISOTROPY and AMPLIMODES program [24, 25]. In a cubic to $R\bar{3}c$ transition, the relevant phonon mode is the a$^-$a$^-$a$^-$ tilting mode with irreducible representation of $R_{4+}$ (shown in Fig. 2(a)). Fig. 2(b) presents the total energy vs. mode amplitude of the tilting, and the angles for relevant tilting are also marked. The total energy of the system is indeed a double-well type potential for the octahedral tilting mode, which means the spontaneous symmetry breaking toward a tilting transition [1]. The energy related to tilting stabilization is quite large (~ 1.66 eV per formula unit for Fig. 2(b)), which means that a transition to a parasitic cubic phase will not occur under ambient conditions, even at high temperatures. The phase transition to a non-centrosymmetric $R3c$ phase is associated with the irreducible representation of $\Gamma_{2-}$, which is the mode of Os-O bending with Li displacement. As shown in Fig. 2(d), the total energy of the system has double well-like shape with a $\Gamma_{2-}$ amplitude, which again indicates the spontaneous breaking of symmetry. The energy related to the $\Gamma_{2-}$ mode was approximately 47 meV per formula unit of perovskite and ~600 K in the temperature scale. This suggests that the experimental transition temperature of 140 K is an order-disorder type with a small (~25%) amount of displacement character. Table 2 shows the results of quantitative analysis of two modes with $R_{4+}$ and $\Gamma_{2-}$ irreducible representation. The reference structure for $R_{4+}$ mode is a cubic structure observed from the $R\bar{3}c$ space group and oxygen tilting is the only component. The mode amplitude (1.6098 Å) is the minimum point in a double well potential, shown in Fig. 2(b), and corresponding tilting angle was 14.11°. The reference structure for the $\Gamma_{2-}$ mode is a rhombohedral structure of the $R\bar{3}c$ space group, and this mode is composed of Os-O bending (Os and O atom displacement) and Li off-centering. The mode amplitude was 0.6010 Å as shown in Fig. 2(d).

Now let us turn our attention to the phonon dispersion calculation. The phonon dispersion curve not only provides a deep understanding of phase transition, but also shows the stability of a non-centrosymmetric phase. The phonon dispersion curve for the two phases was calculated using a frozen phonon approximation. LO/TO splitting was not considered because of the metallic nature of the system. Fig. 3(a) and 3(b) show the phonon dispersion curve along the high symmetry point of Brillouine zone, and the atom projected the partial phonon density of state (PDOS) in the $R\bar{3}c$ phase. The negative y-axis in Fig. 3(a) means the imaginary frequency mode, which is unstable in that structure. The frequency of the zone center phonon is summarized in supplementary Table S1. Two imaginary frequency modes exist at the zone center Γ point, $A_{2g}$ and $A_{2u}$. The irreducible representation of the $A_{2u}$ mode is the $\Gamma_{2-}$ mode in Fig. 2(c) and 2(d). Note that the $A_{2u}$ mode is the soft mode of LiNbO$_3$ [13-18]. The $A_{2u}$ mode is also related to the domain structure in the ferroelectric phase of LiNbO$_3$ to prefer to have domain walls oriented parallel to the threefold symmetry axis [13]. Similar phenomena can be expected in LiOsO$_3$. The atom projected PDOS graph (Fig. 2(d)) suggests that the soft mode is associated with all three atoms. The amplitude of the Os and O atom and symmetry consideration gives the bending of angles with a deviation angle of 178.03°. The deviation angle calculated from the experimental data was approximately 176.84°, as summarized in Table 1. Note that the deviation angles for the three O-Os-O were the same, which is due mainly to off-centering of the Os atom in the [111] pseudo cubic directions. A large contribution of Li atoms can also be observed in this region, indicating the order-disorder nature of the phase transition. On the other hand, for the high frequency optical phonon, the contribution of Li is minimal, suggesting the weakly bounded state of Li in the crystal lattice. The phonon dispersion and atom projected, partial PDOS of the optimized $R3c$ phase in Fig. 3(c) and 3(d) were also checked. The phonon dispersion does not show any imaginary frequency, meaning the stability of this phase [13, 14, 27, 28]. The contribution from a Li atom appears at approximately 6 THz, which is the main difference between the two phases. In the highest optical phonon frequency region, the oxygen contribution is dominant. Importantly, the phonon spectrum of the $R3c$ phase of LiOsO$_3$ is similar to that of ferroelectric LiNbO$_3$ [13-18].

    We are now in a position to discuss about the electronic properties of two phases. Figure 4(a), 4(b), and 4(c) depict the atom projected DOS of LiOsO$_3$ in the two rhombohedra phases near the Fermi level. To check DOS, a careful test was conducted and the final results for DOS were obtained for a dense 20 × 20 × 20 Gamma-centered Monkhorst-Pack k-point grid. Firstly, the main contribution of a Li atom located in the regions ~ 45 eV lower than the Fermi level and the contribution around the Fermi level were 100 times smaller than that of the other two atoms. Note that the scale of Fig. 4(a) is 100 times smaller than that shown in Fig. 4(b) and 4(c). In other words, oxygen and Os hybridization contribute mainly to the Fermi level. Secondly, in LiOsO$_3$, the ionic configuration of Os was 5+, and there were three electrons in the d level. The Os partial DOS of LiOsO$_3$ is similar to that of NaOsO$_3$ and other osnates [29-31]. Thirdly, in the initial view, the atom-projected DOSs of two phases were similar. However, a careful examination reveals that the DOS at the Fermi level for a non-centrosymmetric ferroelectric-like phase is larger. To observe this contribution, the difference between two total DOS (delta DOS is defined by DOS for non-centrosymmetric phase minus DOS for centrosymmetric phase) are plotted in Fig. 4(d). Here, a negative value means that the DOS of the $R\bar{3}c$ centrosymmetric phase is larger than that of the $R3c$ ferroelectric-like phase in a given energy

level, and the summation of delta DOS up to the Fermi level is zero. A negative peak of the delta DOS around -2.5 eV and positive peak centered -0.5 eV from the Fermi level were observed. The delta DOS around the Femi level has a positive value, which indicates that a non-centrosymmetric phase has a larger carrier density around the Fermi level. Because the electronic transport of the metal is determined mainly by $N(E_f)$ (number of carrier at Fermi level), the electronic properties have sharp discontinuity at the phase transition. From the experimental data of the temperature-dependent resistivity, the non-centrosymmetric phase has smaller resistance than the centrosymmetric phase. Higher susceptibility is also expected due to the large number of carriers, and the electronic contribution to heat capacity is larger in the non-centrosymmetric phase. There are clear experimental evidence for the susceptibility and heat capacity enhancement, and the density functional data strongly supports the experimental results [8]. The band structure and Fermi surface topology was also checked (see suppl. Material S1 and S2). The band structures of the two phases were similar with a smaller change near the Fermi level, and the Fermi topology [30, 32] does not show any drastic change between the two phases. Of course, there were small changes in the symmetry of the hole and electronic Fermi surface as well as a smaller change in the cross sectional area of the Fermi surface, indicating that a de Haas van Alphen type experiment will show a small but clear anomaly across the transition. The change in electronic structure might be due to a ferroelectric-like structural change across the phase transition, i.e. the electronic properties were modified due to electron phonon interactions. In addition, the electronic structure calculations suggest that the phonon-related transport properties (such as Hall transport, thermal transport, and angle resolved transport properties) will have a significant change due mainly to two effects: a large change in the phonon structure and a small increase in $N(E_f)$ across the phase transition.

In conclusion, we have studied the phonon and electronic structure of metallic $LiOsO_3$ based on first-principle density functional theory. The non-centrosymmetric phase of $LiOsO_3$, $R\bar{3}c$, is closely associated with phonon mode softening and octahedral tilting, which is similar to the phase transition of ferroelectric $LiNbO_3$-related compounds. The phonon mode of soft phonons ($\Gamma_{2-}$ or $A_{2u}$) and octahedral tilting ($R_{4+}$) was analyzed quantitatively based on density functional theory and group theory. The soft phonon is the main source of the phase transition from the centrosymmetric to non-centrosymmetric phase. The density of states near the Femi level increases across this phase transition, which explains the resistivity, heat capacity, and susceptibility change, due to electron lattice coupling. The density functional calculation provides the overall theoretical background for better understanding the non-centrosymmetric ferroelectric-like transition in metallic $LiOsO_3$. Nevertheless, soft mode experiment (Raman or Neutron Scattering) and transport experiment (such as Hall transport, thermal transport, and angle resolved transport properties) will be needed to achieve a better understanding of a non-centrosymmetric phase transition in this material.

This study was supported by NSF of Korea (NSF-2013R1A1A2004496). The computational resources have been provided by KISTI Supercomputing Center (Project No. KSC-2013-C1-029).

**Figure Captions**

Figure 1. (a) Detailed room temperature structure of LiOsO$_3$ with space group 167, $R\bar{3}c$. Li, Os, and O atoms are drawn as green, grey, and red balls, respectively. The oxygen plane is marked as a red line and [111] direction of a pseudo cubic notation is given as an arrow. (b) Schematic diagram of the oxygen plane vs. Li and Os atoms in the [111] pseudo cubic direction of the $R\bar{3}c$ phase. Os atoms located in the center of two oxygen atoms. (c) Schematic diagram of the ferroelectric-like $R3c$ phase. In this projection, the non-centrosymmetric nature is clearly shown by both Li and Os off centering. (c) The total energy per formula unit as a function of the volume of the perovskite formula unit of cubic, $R\bar{3}c$, and $R3c$ phase.

Figure 2. (a) Octahedral tilting (a$^-$a$^-$a$^-$) observed from the [111] pseudo cubic axis. (b) Total energy variation as a function of the octahedral tiling mode amplitude. The upper axis also shows the tilting angle of the octahedra. (c) $\Gamma_2$- (A$_{2u}$) mode associated with $R\bar{3}c$ to $R3c$ phase transition. (d) Total energy variation as a function of the $\Gamma_2$- (A$_{2u}$) mode amplitude.

Figure 3. Phonon dispersion curve of LiOsO$_3$ with two different phase along the high symmetry point in the Brillouine zone of (a) $R\bar{3}c$ room temperature phase and (c) $R3c$ ferroelectric-like phase. Atom projected phonon partial density of state (pDOS) of (a) $R\bar{3}c$ room temperature phase and (c) $R3c$ ferroelectric-like phase.

Figure 4. Atom-projected electronic density of state (DOS) of (a) Li, (b) Os, and (c) oxygen atom. The blue solid line and the red dashed line denotes the $R\bar{3}c$ room temperature phase and $R3c$ ferroelectric-like phase, respectively. Note on the small y-axis scale of Li atom. (d) The difference between the total DOS of the $R3c$ ferroelectric-like phase and that of $R\bar{3}c$ room temperature phase. The positive value means that the total DOS of the $R3c$ ferroelectric-like phase is larger than that of the $R\bar{3}c$ at a given energy.

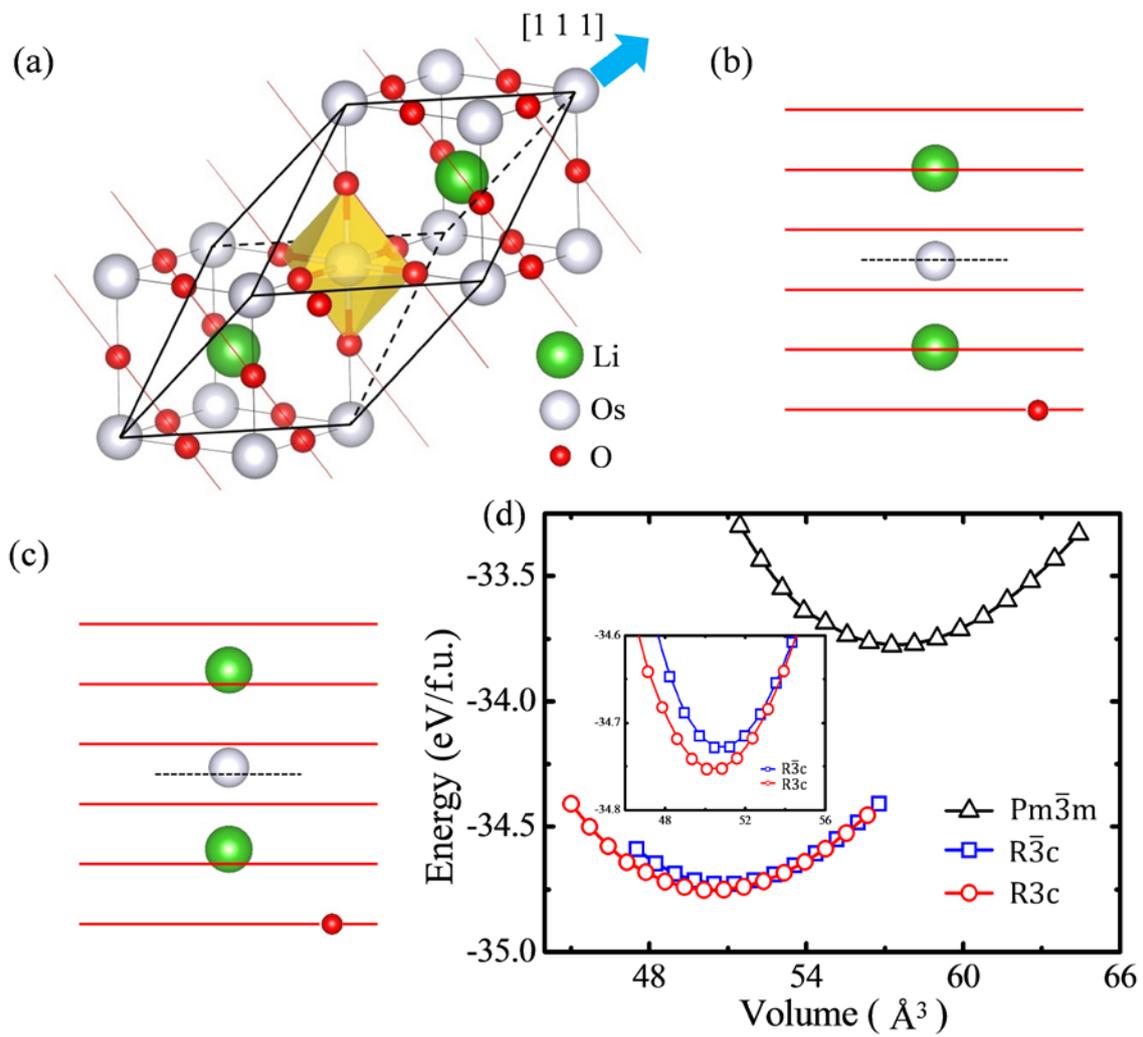

Figure 1 (Color Online) Sim and Kim

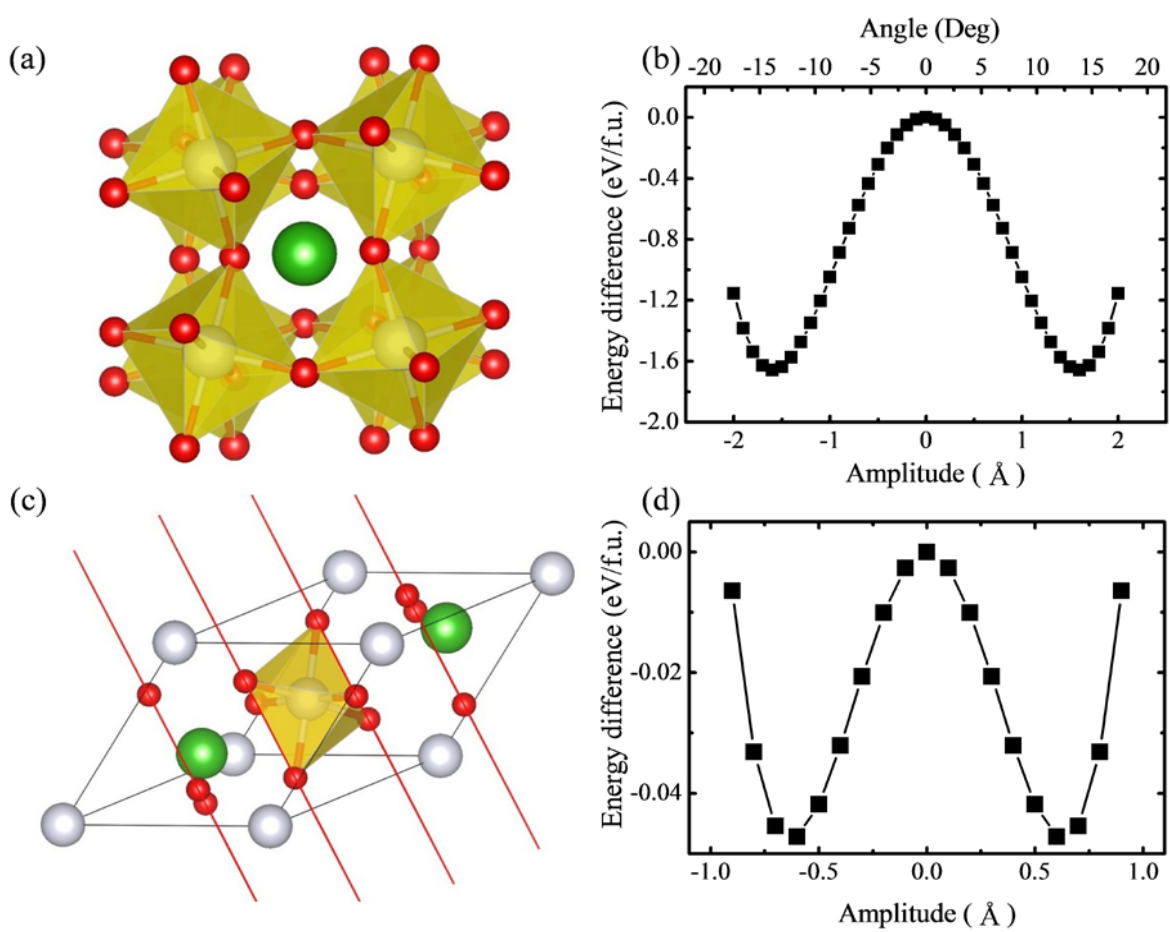

Figure 2 (Color Online) Sim and Kim

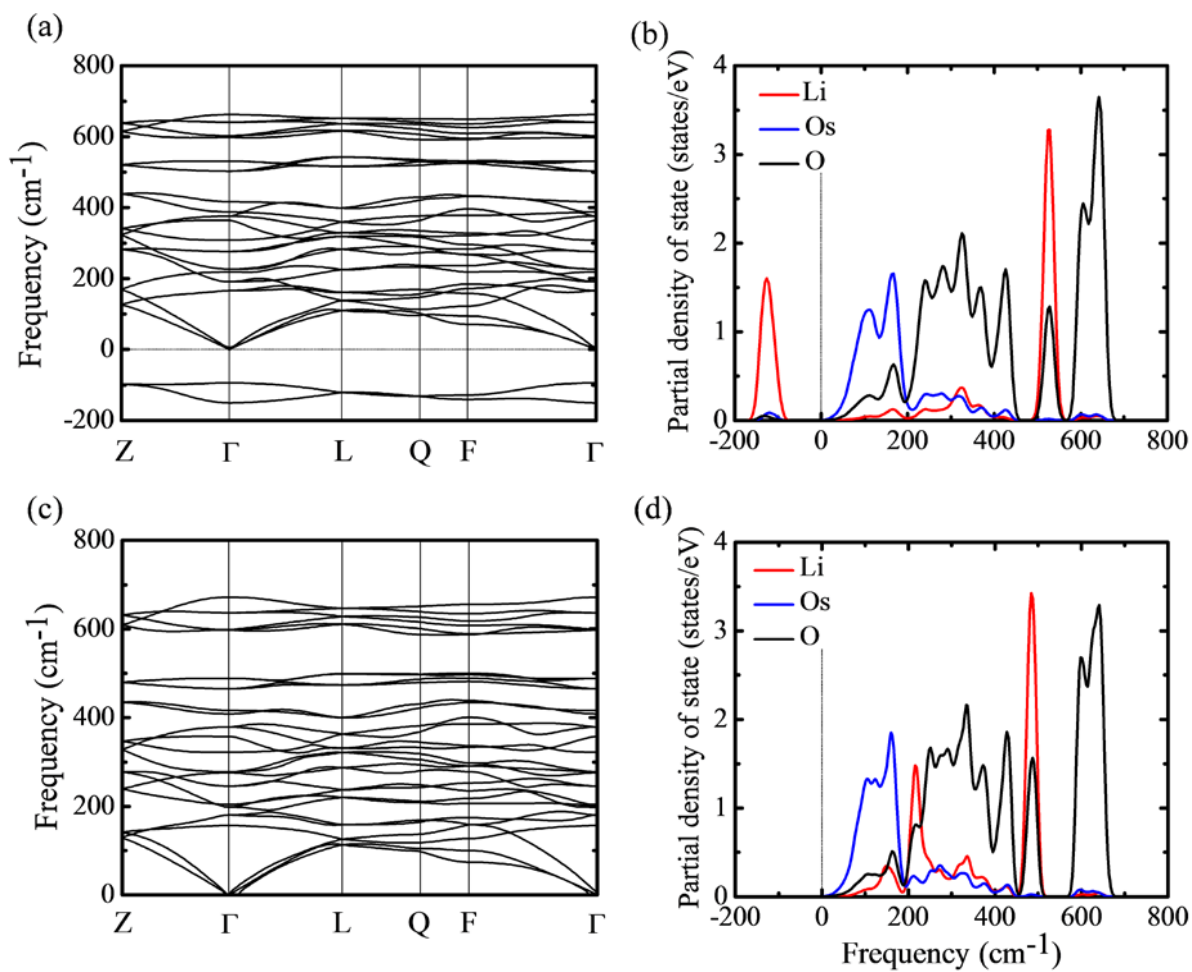

Figure 3 (Color Online) Sim and Kim

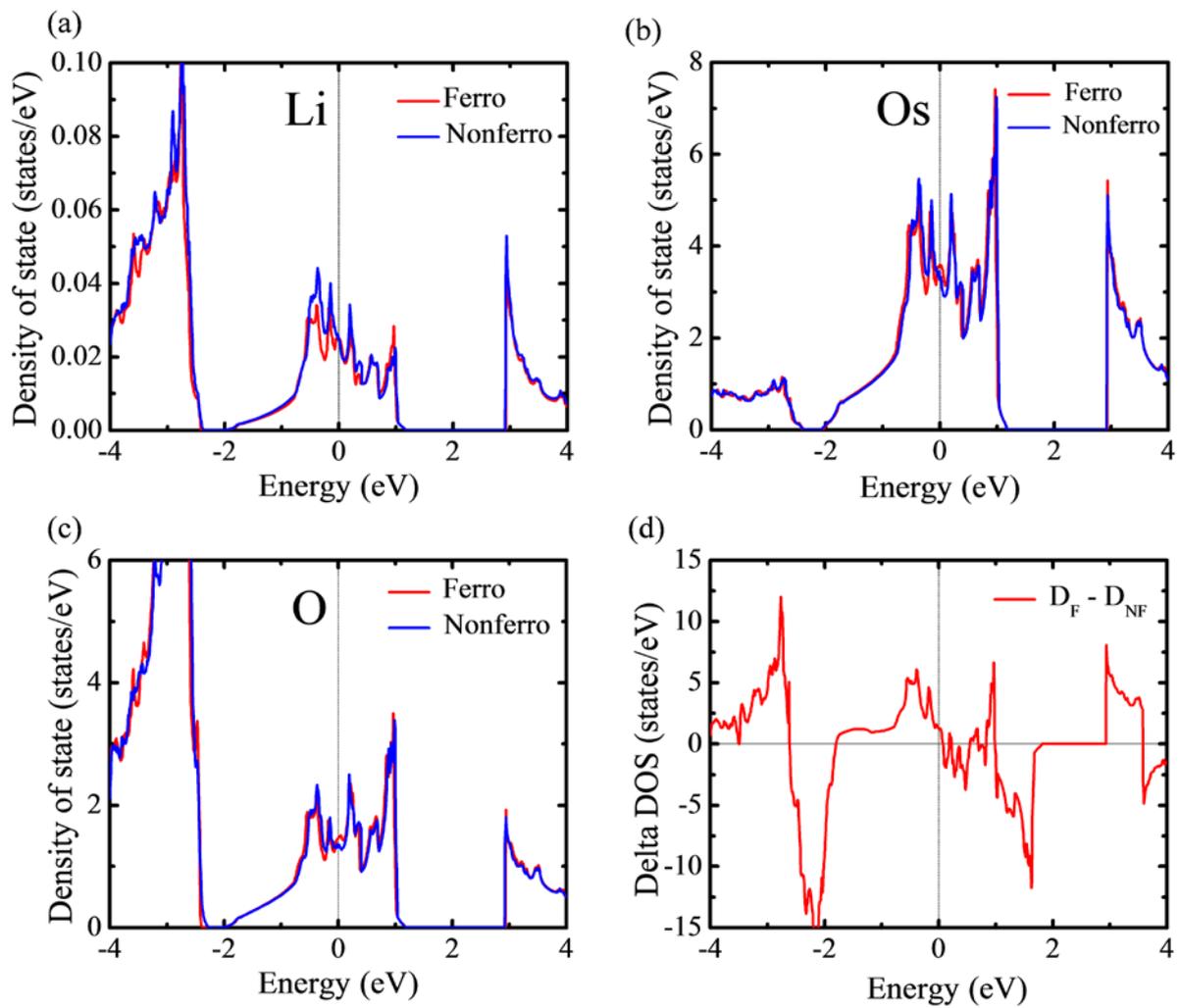

Figure 4 (Color Online) Sim and Kim

TABLE I. Structural parameters and atomic Wyckoff position of LiOsO3 for the two different phases ($R\bar{3}c$ and $R3c$ phase). Here the volume is given for standard conventional cell containing 6 pervoskite formula unit and angle is given for O-Os-O bending angle. The experimental values from ref. 8 are also given.

| LiOsO3 | | a (Å) | c (Å) | Volume (Å³) | Angle (Deg) | Atom | Wyckoff site | x | y | z |
|---|---|---|---|---|---|---|---|---|---|---|
| $R\bar{3}c$ | Cal. | 5.145 | 13.210 | 100.946 | 180 | Li | 6a | 0.0 | 0.0 | 0.25 |
| | | | | | | Os | 6b | 0.0 | 0.0 | 0.0 |
| | | | | | | O | 18e | 0.6278 | 0.0 | 0.25 |
| | Exp. | 5.064 | 13.211 | 97.790 | 180 | Li | 6a | 0.0 | 0.0 | 0.25 |
| | | | | | | Os | 6b | 0.0 | 0.0 | 0.0 |
| | | | | | | O | 18e | 0.6298 | 0.0 | 0.25 |
| $R3c$ | Cal. | 5.093 | 13.378 | 100.178 | 178.03 | Li | 6a | 0.0 | 0.0 | 0.2164 |
| | | | | | | Os | 6a | 0.0 | 0.0 | 0.0 |
| | | | | | | O | 18b | 0.6296 | -0.0339 | 0.2523 |
| | Exp. | 5.046 | 13.239 | 97.293 | 176.84 | Li | 6a | 0.0 | 0.0 | 0.2147 |
| | | | | | | Os | 6a | 0.0 | 0.0 | 0 |
| | | | | | | O | 18b | 0.6260 | -0.0102 | 0.2525 |

TABLE II. Amplimode result of LiOsO3 for the two different phases ($R\bar{3}c$ and $R3c$ phase). K-vector, character, and amplitude of each mode are given. $R_{4+}$ is octahedral tilting mode responsible for cubic to $R\bar{3}c$ phase transition and $\Gamma_{2-}$ ( or $A_{2u}$) is soft phonon mode associated with $R\bar{3}c$ to ferroelectric like $R3c$ phase.

| LiOsO3 | | Reference structure | | | | Displacement | | |
|---|---|---|---|---|---|---|---|---|
| | Atom | wyckoff | x | y | z | dx | dy | dz |
| $R_{4+}$ (from cubic) | Li | 6a | 0.0 | 0.0 | 0.25 | 0.0 | 0.0 | 0.0 |
| | Os | 6b | 0.0 | 0.0 | 0.0 | 0.0 | 0.0 | 0.0 |
| | O | 18e | 0.1667 | 0.3333 | 0.0833 | 0.0794 | 0.0 | 0.0 |
| $\Gamma_{2-}$ (from $R\bar{3}c$) | Li | 6a | 0.0 | 0.0 | 0.25 | 0.0 | 0.0 | 0.0471 |
| | Os | 6a | 0.0 | 0.0 | 0.5 | 0.0 | 0.0 | -0.0088 |
| | O | 18b | 0.3688 | 0.3688 | 0.25 | 0.0027 | -0.0027 | -0.0128 |

## Supplementary information

Hyunsu Sim and Bog G. Kim

*Department of Physics, Pusan National University, Pusan, 609-735, South Korea*

Some of detailed calculation results are summarized here. Other results can be easily obtainable using the parameters presented in the manuscript. Input and Output files for some calculations can be provided upon e-mail request (boggikim@pusan.ac.kr).

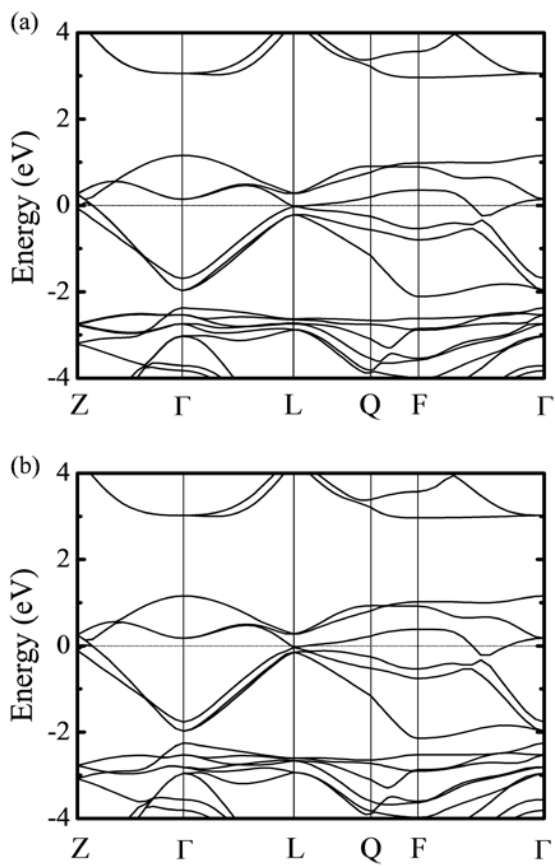

Figure S1. Electronic band structure along high symmetry point in Brillouine zone of (a) $R\bar{3}c$ room temperature phase and (b) $R3c$ ferroelectric-like phase.

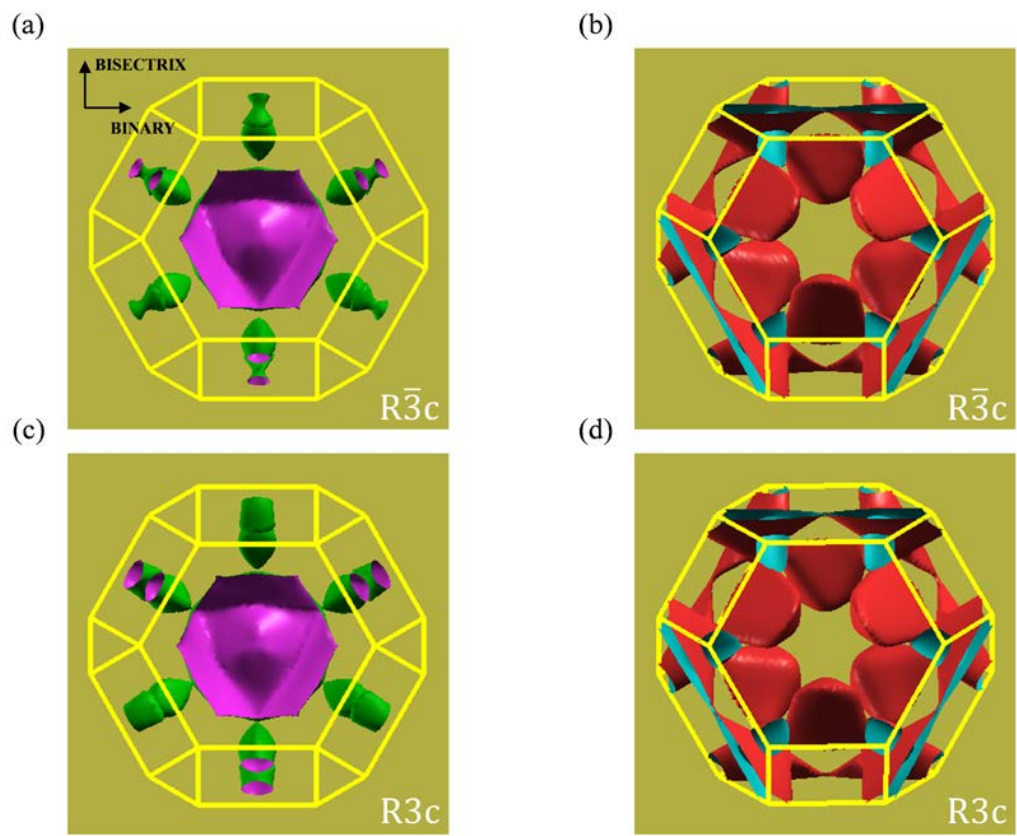

Figure S2. (a) Hole and (b) electron Fermi surface topology of the $R\bar{3}c$ room temperature phase. (c) Hole and (d) electron) Fermi surface topology the $R3c$ ferroelectric-like phase.

|  | R3c | | R$\bar{3}$c | |
| --- | --- | --- | --- | --- |
| No. | Irred. Rep. | Frequency (THz) | Irred. Rep. | Frequency (THz) |
| 1. | $A_2$ | 4.751 | $A_{2u}$ | -4.581 |
| 2. | E | 5.207 | $A_{2g}$ | -3.049 |
| 3. | E | 6.012 | $E_u$ | 5.142 |
| 4. | $A_1$ | 6.327 | $E_g$ | 5.977 |
| 5. | E | 7.344 | $A_{1u}$ | 6.780 |
| 6. | $A_2$ | 8.422 | $E_u$ | 6.873 |
| 7. | E | 8.451 | $E_u$ | 8.463 |
| 8. | $A_1$ | 9.841 | $A_{2u}$ | 9.453 |
| 9. | $A_2$ | 10.787 | $A_{1u}$ | 11.109 |
| 10. | E | 11.560 | $E_g$ | 11.496 |
| 11. | $A_2$ | 12.393 | $A_{2u}$ | 11.820 |
| 12. | $A_1$ | 12.766 | $A_{1g}$ | 12.599 |
| 13. | E | 14.137 | $E_u$ | 15.210 |
| 14. | E | 14.852 | $E_g$ | 16.070 |
| 15. | $A_1$ | 18.279 | $A_{2u}$ | 18.140 |
| 16. | E | 18.332 | $E_g$ | 18.263 |
| 17. | E | 19.433 | $E_u$ | 19.441 |
| 18. | $A_2$ | 19.849 | $A_{2g}$ | 20.103 |

Table S1. Optical phonon frequency (in THz) of Γ point in the R$\bar{3}$c phase and R3c ferroelectric-like phase of LiOsO$_3$. At Γ point in the R$\bar{3}$c phase, the optical phonon can be classified according to the irreducible representations into $A_{1g} \oplus 2A_{1u} \oplus 3A_{2g} \oplus 3A_{2u} \oplus 4E_g \oplus 5E_u$, where $A_{2u}$ and $E_u$ modes are infrared active. At Γ point in the R3c phase, the optical phonon can be classified according to the irreducible representations into $4A_1 \oplus 5A_2 \oplus 9E$, where $A_1$ and E modes are Raman and Infrared active. Imaginary value ($A_{2g}$ and $A_{2u}$) means unstable phonon and $A_{2u}$ mode is soft mode associated with ferroelectric like phase transition.